# Photonic hooks from Janus microcylinders

GUOQIANG GUO,[1,2] LIYANG SHAO,[1,*] JUN SONG,[2,5] JUNLE QU,[2] KAI ZHENG,[2] XINGLIANG SHEN,[1] ZENG PENG,[1] JIE HU,[1] XIAOLONG CHEN,[1] MING CHEN,[3] AND QIANG WU[4]

[1]*Department of Electrical and Electronic Engineering, Southern University of Science and Technology, Shenzhen 518055, China*
[2]*Key Laboratory of Optoelectronic Devices and Systems of Ministry of Education and Guangdong Province, College of Physics and Optoelectronic Engineering, Shenzhen University, Shenzhen 518060, China*
[3]*Center for Information Photonics and Energy Materials, Shenzhen Institutes of Advanced Technology, Chinese Academy of Sciences, Shenzhen 518055, China*
[4]*Department of Mathematics, Physics and Electrical Engineering, Northumbria University, Newcastle Upon Tyne, NE1 8ST, United Kingdom*
[5]*songjun@szu.edu.cn*
*\*shaoly@sustech.edu.cn*

**Abstract:** Recently, a type of curved light beams, photonic hooks (PHs), was theoretically predicted and experimentally observed. The production of photonic hook (PH) is due to the breaking of structural symmetry of a plane-wave illuminated microparticle. Herein, we presented and implemented a new approach, of utilizing the symmetry-broken of the microparticles in material composition, for the generation of PHs from Janus microcylinders. Finite element method based numerical simulation and energy flow diagram represented theoretical analysis were used to investigate the field distribution characteristics and formation mechanism of the PHs. The full width at half-maximum (FWHM) of the PH (~0.29$\lambda$) is smaller than the FWHM of the photonic nanojet (~0.35$\lambda$) formed from a circular microcylinder with the same geometric radius. By changing the refractive index contrasts between upper and lower half-cylinders, or rotating the Janus microcylinder relative to the central axis, the shape profiles of the PHs can be efficiently modulated. The tunability of the PHs through simple stretching or compression operations, for the Janus microcylinder constituted by one solid inorganic half-cylinder and the other flexible polymer half-cylinder, was studied and discussed as well.

## 1. Introduction

A fundamental point in geometric optics is that light travels in straight lines in a homogeneous medium [1]. Such common sense was first broken by the experimental observation of Airy beams propagating along a curved parabolic trajectory in space [2]. The propagated wave-packet of Airy beams is essentially a solution to the Schrödinger equation [3]. From the physical point of view, the parabolic-shaped field distribution, is a result of the superposition of many rays in which the main lobe representing the family of the rays asymptotically approaching beam caustic [4]. This makes the Airy beams showing unusual properties (nondiffracting, self-healing and accelerating) and possessing tremendous potential in many fields (e.g., optical mediated particle clearing, curved plasma channel generation, light bullet generation, optical trapping, electron acceleration, super-resolution imaging, etc.) [5]. Last year, a second new class of curved light beam, the so-called "photonic hook" (PH), was formally put forward [6,7]. Unlike the Airy beam, the emerging PH is formed from a plane-wave-illuminated dielectric microparticle of which is composed of a cuboid and a triangular prism. The generation process is simpler relative to the formation of Airy beams. The startling qualities of sub-diffraction-limited beam waists and sub-wavelength-scaled curvature radius show important application prospects in the fields of optical imaging, nanoparticle manipulation, cell redistribution, nonlinear optics, integrated optics, and so on [8]. The unique application of generating optical force for transporting nanoparticle along a curvilinear trajectory is especially attractive for sub-wavelength optical micromanipulation in physics and cell-biology, and lab-on-a-chip imaging in microfluidic engineering.

PH was later experimentally verified in the terahertz range [9], and further extended to the fields of surface plasmons and acoustic waves [10,11]. This intriguing phenomenon, in one sense, can be regarded as a curved photonic jet (PJ). As is well known, photonic jets (PJs) are generally obtained by illuminating the dielectric microparticles of two-dimensional (2D) axial symmetries or three-dimensional (3D) spherical symmetries [12-16]. The shape features and field distributions of the PJs, under given conditions of certain environmental medium and illumination source, are depended on the geometrical morphologies and material properties of the investigated dielectric microparticles [17-20]. By introducing a triangular prism adjacent to the front side of a cuboid surface [6,7], the phase velocity and wave interference of the transmitted light in the upper and lower parts of the microparticle are different due to the break of particle's shape symmetry, and thus producing a curved PJ, i.e. PH. Along a similar line of thought, it is natural to think that there should exist another way to form photonic hooks (PHs) by breaking the symmetry of the microparticles in material composition. The particles, named after the two-faced Roman God Janus, provide the possibilities for the realization of attaining PHs in this way [21]. Such particles can dispose particle architectures in an asymmetric fashion and impart different chemical or physical properties on both sides of the particles [22,23].

In this paper, we proposed and demonstrated another new approach for the generation of PHs from Janus microcylinders. This kind of microparticles are constituted by two micro-sized half-cylinders of the same shapes, but the different dielectric properties. Relevant studies were conducted through a finite element method (FEM) based 2D numerical simulation model. With respect to the traditional PJs from single dielectric microcylinders, PHs not only show obvious bending characteristics but also have smaller full width at half-maximum (FWHM). The theoretical analysis results on the basis of energy flow distributions and the corresponding Poynting vector fields illustrate the generation mechanism of PHs from Janus microparticles. By varying the refractive index (RI) contrasts of the two half-cylinders, the inflection points and bending degrees of the PHs, to some extent, can be effectively modulated. When rotating the Janus microcylinder in a counterclockwise or clockwise direction, the "hook angles" of which representing the curvatures of PHs change with the variations of rotation angles in a circular fashion. The tunability of PHs, generated from a kind of Janus microcylinder composed of one solid inorganic half-cylinder and the other flexible polymer half-cylinder, was considered and discussed by simply exerting mechanical stretching or compression onto the polymer half-cylinder.

## 2. Research approach

Two main computational techniques of FEM and finite difference time-domain method are widely applied to investigate the optical field distribution inside and outside the dielectric microparticle illuminated by a plane-wave [24]. In our work, FEM-based COMSOL Multiphysics commercial software package was adopted to carry out the 2D full-wave simulations.

Figures 1(a) and 1(b) are the schematic diagrams showing the 3D stereogram and 2D sectional view of the investigated model. The refractive indices (RIs) of the two half-cylinders are respective $n_2$ and $n_3$, and the whole Janus microcylinder is placed in the air ($n_1 = 1$). The incident light beam is a monochromatic plane-wave with a wavelength of $\lambda = 532$ nm, light intensity of $I = 1$ and propagation direction along the transmission axis. For the entire computational domain, non-uniform meshes with RI-dependent element size were used here. The perfectly matched layer absorbing boundary condition was set, on the one end side and two lateral sides of the computational domain, to fully absorb the outgoing waves and avoid undesired back reflections.

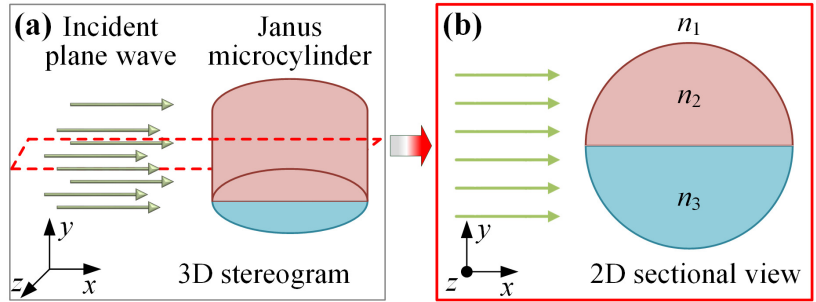

Fig. 1. Schematics of plane-wave-illuminated Janus microcylinder: (a) 3D stereogram and (b) 2D sectional view.

## 3. Results and discussion

In general, the optical fields on the rear side of the single dielectric microcylinder, as shown in Fig. 2(a), are symmetrically distributed on both sides of the symmetrical axis in the direction of $x$-axis which is the same as the research results in Ref. 25. The maximum light intensity ($I_{max}$) is on the axis of symmetry [red dot marked in Fig. 2(a)]. When the single dielectric microcylinder was replaced by a Janus microcylinder, as can be seen from the top and bottom half of the constituted microcylinder in Fig. 2(b), the position of $I_{max}$ deviated from the axis of symmetry and the focusing light beam was clearly bent to form a PH on the back surface of the microparticle. Here, the RI of one half-cylinder [lower half-cylinder in Fig. 2(b)] is $n_3 = 1.4$, while the RI of the other half-cylinder [upper half-cylinder in Fig. 2(b)] is $n_2 = 1.46$ which is the same as the RI of microcylinder given in Fig. 2(a). The diameters of the microcylinders in both two cases were set as four times the wavelength $\lambda$.

Figure 2(c) shows the lateral optical field distributions of the produced PJ and PH in the points of $I_{max}$. It is observed that the FWHM of PH (~$0.29\lambda$) is far below the diffraction limit and less than the FWHM of PJ (~$0.35\lambda$) as well. The lines connected by the specific points on the right end of each contour line, which are marked in green and red dotted lines in Figs. 2(a) and 2(b), clearly show the intensity distributions along a straight trajectory for PJ and a curve trajectory for PH. This situation can also be seen from the 2D intensity distributions in Fig. 2(d). All the points representing the green dotted line are on the $x$-axis, while the points on the red dotted line move away from the $x$-axis in the beginning and then slowly approach towards the $x$-axis. With regard to $I_{max}$, the value for PH is ~10.2 which is a little less than the value of ~10.9 for the case of PJ. The reason is probably because some total internal reflection occurs near the output end of Janus microcylinder. When the light interacts with the interface between upper and lower half-cylinders, there is a small fraction of the light propagating from optically dense medium ($n_2 = 1.46$) into optically thinner medium ($n_3 = 1.4$) and thus causing total internal reflection. This little bit of light, for the focused light beam on the rear side of the Janus microcylinder, is equivalent to loss in the ways of optical reflection or scattering. As can be seen from the comparison of optical field distributions around the region of light output end in Figs. 2(a) and 2(b), the bright areas in the upper half-cylinder are obviously larger than the bright areas in the case of single dielectric microcylinder.

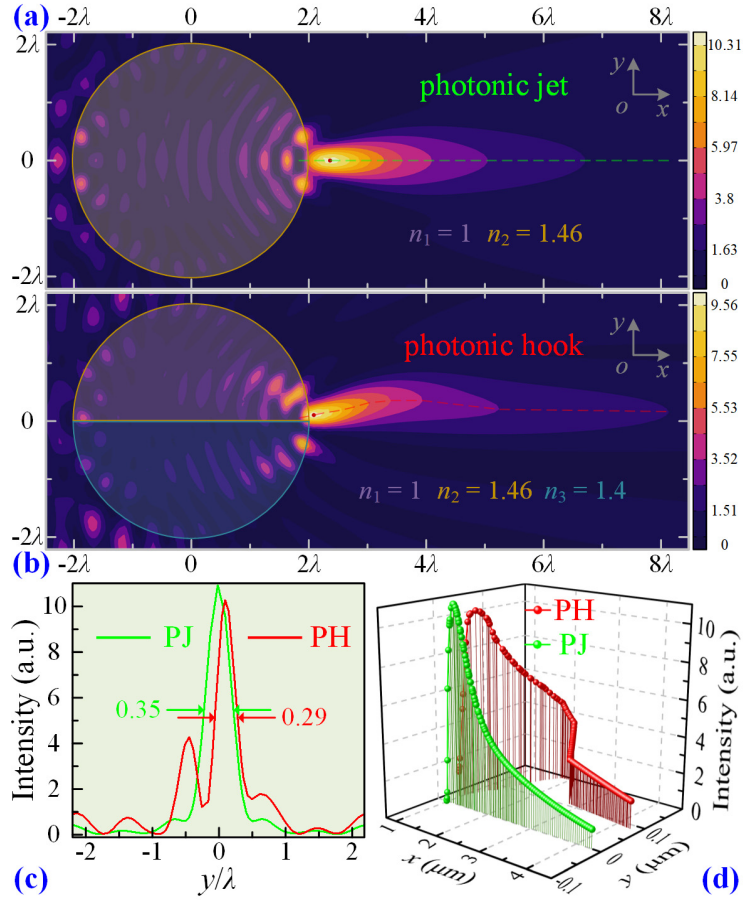

Fig. 2. (a) PJ and (b) PH formed by illuminating a single dielectric microcylinder (diameter: $4\lambda$, RI: $n_2 = 1.46$) and a Janus microcylinder (diameter: $4\lambda$, RI: $n_2 = 1.46$, $n_3 = 1.4$). (c) The optical field profiles at $I_{max}$ points along $y$ direction for PJ (green curve) and PH (red curve). (d) The optical field distributions of the trajectories marked with green dotted line and red dotted line in (a) and (b). PJ: photonic jet, PH: photonic hook.

According to classical electromagnetic theory and our previous research, the energy flow diagram denoted by the field lines of time-averaged Poynting vectors can be used to compare and analyze the formation mechanism of PJs and PHs [26,27]. As shown in Fig. 3(a), the Poynting vector fields (conical arrows), in the case of conventional microcylinder, are always symmetric relative to the axis of symmetry (black dotted line) in the whole computational domain. On the rear side of the microcylinder, the Poynting vector fields and energy flow lines within the auxiliary green-dotted-line converged into a focused beam of typical PJ. While in Fig. 3(b), the intensity of Poynting vector fields in the region near the emergent points and the configuration of energy flow streamlines on the back surface are respective higher and denser above the $x$-axis. The difference of dielectric properties between two half-cylinders resulted in the broken of microparticle's symmetry in material composition and eventually led to the formation of curved beam profile in the case of Janus microcylinder. For the 2D full-wave simulation model in this work, the electrical and magnetic vector flux is on the $x$-$y$ plane.

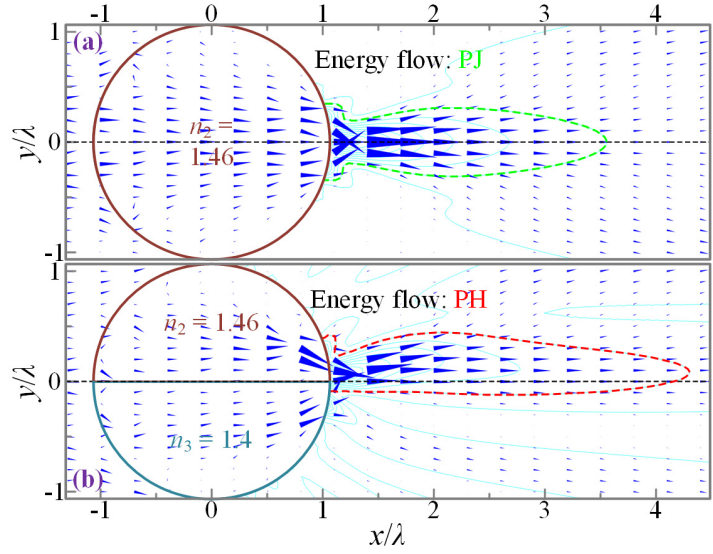

Fig. 3. Poynting vector distributions (blue conical arrows) and energy flow streamlines (cyan curves) for (a) conventional microcylinder and (b) Janus microcylinder with diameter of $4\lambda$ in the $x$-$y$ plane.

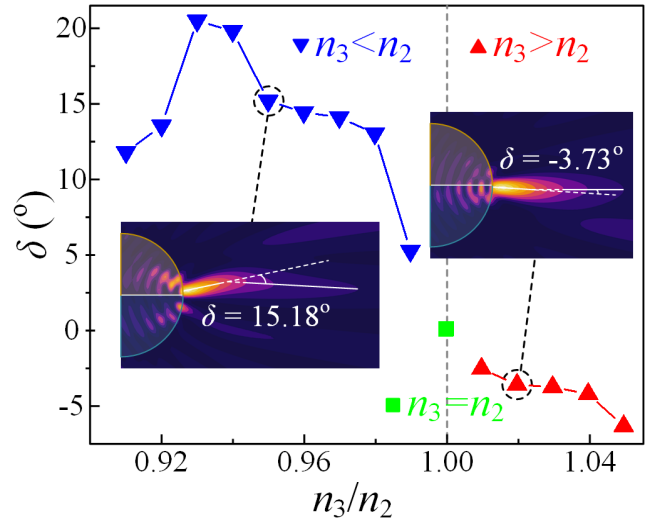

Fig. 4. Bending angle $\delta$ with RI contrasts $n_3/n_2$ from 0.91 to 1.05. Insets: the electrical field distributions and marked bending angles for the cases of $n_3/n_2$ equaling 0.95 and 1.02.

Referring to the practice in Ref. 7, we studied the variations of the bending angle ($\delta$) with RI contrast ($n_3/n_2$) for illustrating the curvatures of the PHs. When the connecting line between the start point and the inflection point turned clockwise to the connecting line between the inflection point and the end point, the bending angle $\delta$ was defined as positive. Conversely, when the deflection was in the opposite direction, $\delta$ was defined as negative. The RI of lower half-cylinder was fixed as $n_2 = 1.46$, the RIs of upper half-cylinder were changing from $0.91n_2$ to $1.05n_2$. As shown in Fig. 4, the bending angles $\delta$ first increase and then decrease as the ratio

$n_3/n_2$ increases from 0.91 to 0.99. All the bending angles are greater than zero and the curved strong field areas are all located above the *x*-axis. The illustration shows $\delta = 15.18°$ in the case of $n_3 = 0.95n_2$. The maximum bending angle is over 20 degrees around $n_3/n_2 \sim 0.93$. If the ratio $n_3/n_2$ increases to 1, the PH turns into PJ due to the Janus microcylinder recovering its symmetry of material compositions for two half-cylinders. When $n_3/n_2 > 1$, the focused light beams began to bending below the *x*-axis and reforming PHs with negative $\delta$. The inset in the right panel of Fig. 4 shows the bending angle $\delta$ equaling -3.73° at $n_3 = 1.02n_2$. Obviously, researches in this area indicate that the degree and direction of deflections for PHs are connected with the RI contrasts between upper and lower half-cylinders of the investigated Janus microcylinder.

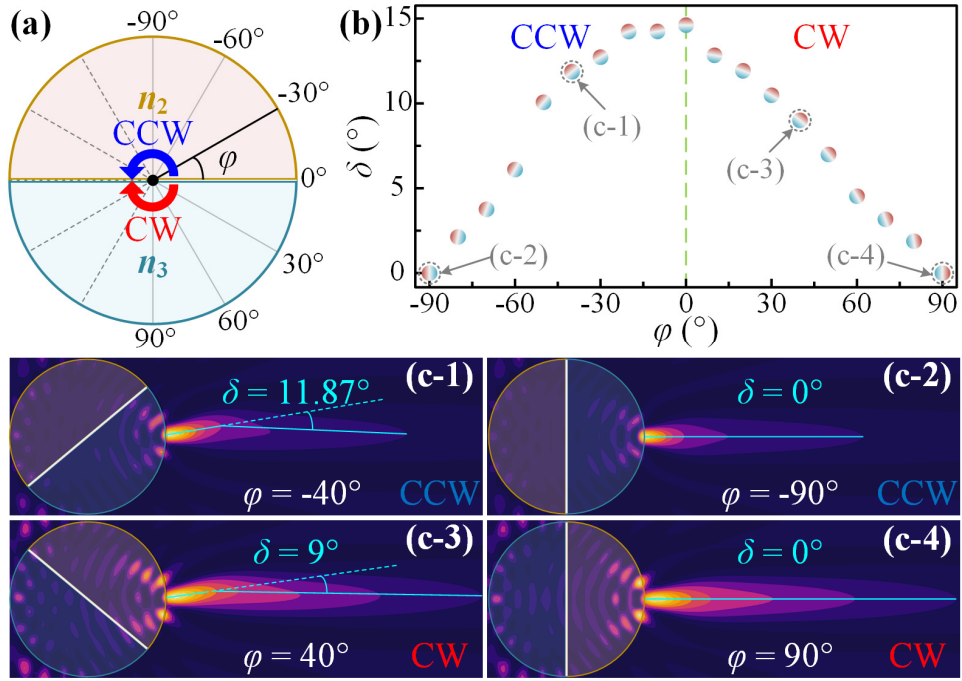

Fig. 5. (a) Clockwise or counterclockwise rotation of plane-wave-illuminated Janus microcylinder with $n_2 = 1.46$ and $n_3 = 1.4$. $\varphi$: rotation angle, CW: clockwise, CCW: counterclockwise. (b) Bending angle $\delta$ as a function of rotation angle $\varphi$. (c) Electric field distributions and bending angles $\delta$ for $\varphi = -40°, -90°, 40°$ and $90°$, selected in (b).

The tunability of PH's bending characteristics is significant for moving objects along specific curved trajectories in the areas of optical trapping or manipulating nanoparticles, biomolecules, cells and viruses. Benefitting from the properties of infinite length in one direction and structural symmetry of Janus microcylinder, it is easy to rotate such microparticle around the central axis of symmetry. In this way, the incident plane-waves hit the boundary profiles of the half-cylinders of which involving two different dielectric materials with different component ratios, and form PHs of different shapes and curvatures. As shown in Fig. 5(a), the Janus microcylinder was rotated clockwise (red arrow) and counterclockwise (blue arrow) about the transmission axis, respectively. The material properties of Janus microcylinder, the surrounding medium and the illumination source are the same with Fig. 2. Here, the geometric center of the structure was defined as the origin, $\varphi$ as the rotation angle where the value obtained by rotating clockwise is positive and negative in the opposite direction. In view of the axial symmetry in geometry of a Janus microcylinder, the rotation angle ranging from -90° to 90° was considered. The step size of rotation angle for each adjustment is 10°. As can be seen from Fig. 5(b), the maximum bending angle ($\delta$ = 14.6°) of the PHs happens at the rotation angle of $\varphi$ =0°. With the increase $\varphi$, no matter in a clockwise or counterclockwise direction, the bending angles $\varphi$ are both getting smaller. Figure 5(c-1) to 5(c-4) show the electric field distributions and the calculated bending angles for the following four cases: $\varphi$ = -40°, -90°, 40° and 90°. The values of bending angles, which are also corresponding to the indicated four cases in Fig. 5(b), are $\delta$ = 11.87°, 0°, 9° and 0°. If we extend the parts of the PHs that are not shown, the bending angles $\delta$ can be circularly restructured back and forth within ±15°.

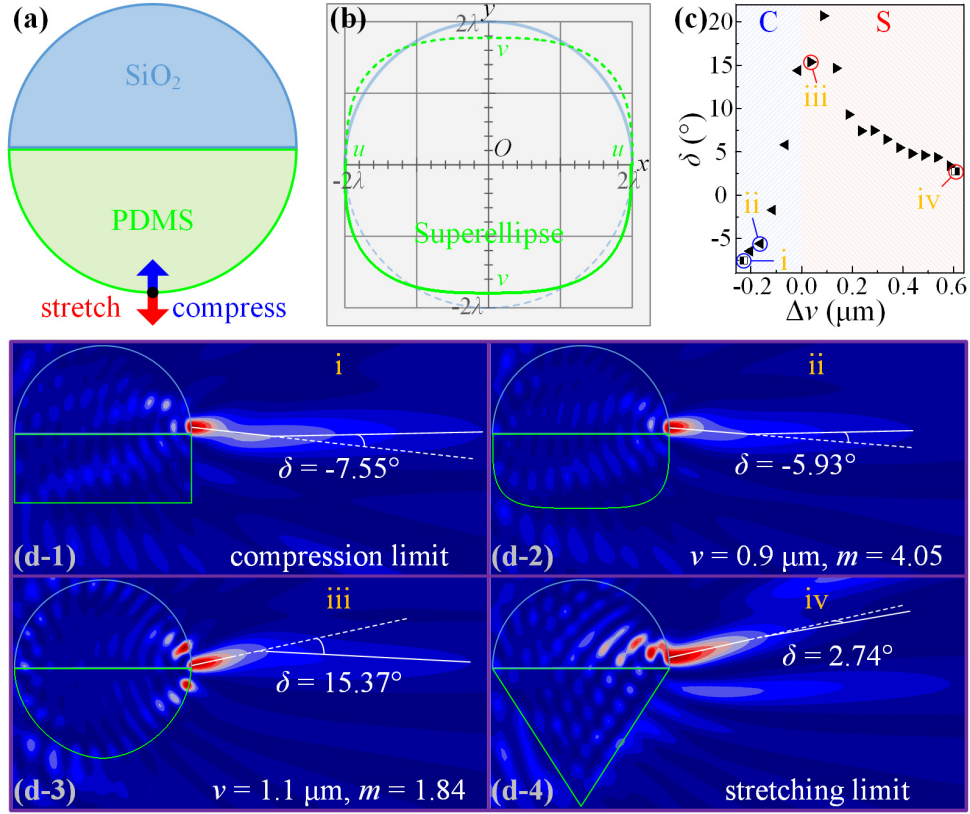

Fig. 6. (a) Stretching and compression tuning of PHs from a Janus microcylinder constituted by solid $SiO_2$ half-cylinder and flexible PDMS half-cylinder. (b) Profile curves of Janus microcylinder before (solid and dashed blue-grey lines) and after (solid blue-grey and green lines) mechanical stretch or compression. The dashed green line is the other half of superellipse curve. $u$: semi-major axis length, $v$: semi-minor axis length. (c) Bending angle $\delta$ of PH as a function of deformation amount $\Delta v$. (d) Electric field distributions and bending angles $\delta$ for four cases: (d-1) compression limit, (d-2) $v$ = 0.9 μm, $m$ = 4.05, (d-3) $v$ = 1.1 μm, $m$ = 1.84, and (d-4) stretching limit.

If one half-cylinder of the Janus microcylinder is an elastic flexible material, the resultant PHs can be adjusted by another method of mechanical stretch or compression. Such tuning mode is discussed with numerical experiments in the following content. Figure 6(a) illustrates the schematic of the stretching and compression tuning. The polydimethylsiloxane (PDMS) was chosen as the stretchable half-cylinder because of its optical transparency and good viscoelasticity [28]. Solid silica was adopted as the other fixed half-cylinder. As shown in Fig. 6(b), the shape profiles were assumed to remain constant for silica half-cylinder (solid blue-grey line) and deform in a superellipse way for PDMS half-cylinder (solid green line) during the process of stretching or compression. The curves indicated by dashed blue-grey line and dashed green line are the other halves of the circle and superellipse, respectively. In the Cartesian coordinate system, the origin $O$ is located at the geometric center of the Janus microcylinder, the radius of the circle is $2\lambda$. The set of all points $(x, y)$ on the superellipse satisfies the equation [29]:

$$\left|\frac{x}{u}\right|^m + \left|\frac{y}{v}\right|^m = 1 \tag{1}$$

where $u$ and $v$ are the lengths of the semi-major and semi-minor axes, $m$ represents the exponent. The exponent $m$ corresponding to each value of $v$ can be solved according to the principle of constant area before and after stretching or compression. The area inside the superellipse curve is expressed in terms of the gamma function [29]:

$$Area = 4uv \frac{\left(\Gamma\left(1+\frac{1}{m}\right)\right)^2}{\Gamma\left(1+\frac{2}{m}\right)} \tag{2}$$

Through the above two formulas, the profile curves resulting from every stretchable action can be figured out. Figure 6(c) shows the relationship between bending angles and amount of stretching or compression. The shapes of the PDMS section in the compression and stretching limits, which are marked as "i" and "iv" in Fig. 6(c) and shown with filled contour plot in Figs. 6(d-1) and 6(d-4), are rectangle and triangle with the diameter of upper half-cylinder as base. The corresponding bending angles are respective $\delta$ = -7.55° and $\delta$ = 2.74°. In the compression range, the bending angles $\delta$ change from negative to positive with the decrease of compression amount $\Delta v$. While the bending angles, in the range of stretching ($\Delta v > 0$), are all deflected clockwise. The maximum bending angle is 20.7°. Two examples, for the cases of compression and stretching, with $v$ = 0.9 μm, $m$ = 4.05 and $v$ = 1.1 μm, $m$ = 1.84 are illustrated in Figs. 6(d-2) and 6(d-3). The simple stretching and compression tunings provide a new way for the manipulation and controlling of PHs.

## 4. Conclusions

In conclusion, the formation of PHs from Janus microcylinders was investigated in detail in this study. Such microcylinders are constituted by two half-cylinders of which having the same geometrical shapes and the different material properties. The results from FEM simulation and Poynting vector analysis prove that the FWHM of PH is smaller than the analogue of PJ and far less than the diffraction limit. Three ways of changing the RI contrasts of the two half-cylinders, rotating the Janus microcylinder relative to the center axis, or exerting mechanical stretching/compression actions onto flexible polymer half-cylinder can be used to flexibly and efficiently adjust the curvatures, deformations and profiles of the PHs. This work provides a new way for the producing of PHs which will promote greater understanding to the formation mechanism and beam properties of PHs.

**Funding**

Shenzhen Postdoctoral Research Grant Program (K19237504); Startup Fund from Southern University of Science and Technology and Shenzhen government (Y01236228 & Y01236128).

**Disclosures**

The authors declare that there are no conflicts of interest related to this article.